\documentstyle[12pt,aasms]{article}

\received{10 April 1995}
\slugcomment{Submitted to the Astronomical Journal}

\begin{document}

\newcommand{\HI}{H~{\small I} }
\newcommand{\HII}{H~{\small II} }
\newcommand{\HH}{H$_{2}$ }
\newcommand{\HA}{H$\alpha$ }

\title{
DISTRIBUTION OF BLUE GALAXIES
IN A MERGING CLUSTER OF GALAXIES ABELL 168
\altaffilmark{1}
}

\author{
A{\small KIHIKO}~T{\small OMITA},
F{\small UMITAKA}~E.~N{\small AKAMURA},
T{\small ADAFUMI}~T{\small AKATA}\altaffilmark{2},
K{\small OUICHIROU}~N{\small AKANISHI},
T{\small SUTOMU}~T{\small AKEUCHI}, {\small AND}
K{\small OUJI}~O{\small HTA}
}
\affil{
Department of Astronomy,
Faculty of Science, 
Kyoto University,
\\
Sakyo-ku,
Kyoto~606-01,~Japan
\\
Electronic mail:
atomita,
fuming,
takata,
nakanisi,
takeuchi,
ohta@kusastro.kyoto-u.ac.jp
}

\and

\author{
T{\small ORU}~Y{\small AMADA}
}
\affil{
Cosmic Radiation Laboratory,
The Institute of Physical and Chemical Research (RIKEN),
\\
Wako,
Saitama~351-01,~Japan
\\
Electronic mail:
yamada@cricket.riken.go.jp
}

\altaffiltext{1}
{Based on observations made at Kiso Observatory.
Kiso Observatory is operated by Institute of Astronomy,
Faculty of Science, University of Tokyo, Japan.}

\altaffiltext{2}
{Present address:
National Astronomical Observatory,
Mitaka, Tokyo 181, Japan;
Electronic mail:
takata@optik.mtk.nao.ac.jp}

\begin{abstract}

The fraction of star-forming galaxies
in rich clusters of galaxies
increases rapidly with the redshift.
This is interpreted as the result of
a rapid evolution of cluster galaxies,
though its mechanism is not yet clear.
One hypothesis is that
if galaxies run into the dense intracluster medium (ICM) regions,
starbursts can be induced due to compression of molecular clouds
in the galaxies
by a raised ``external'' ICM pressure.
In a merging cluster,
there would be dense ICM regions and
some galaxies may experience
a rapid increase of external pressure. 
Thus,
if the mechanism works,
starbursts would occur in such galaxies.
Such a situation is considered to be realized in Coma cluster,
and galaxies showing a recent starburst have
a characteristic spatial distribution;
they populate an elongated region between two sub-clusters.
We examined the above hypothesis
by measuring the spatial distribution of blue galaxies,
regarded as star-forming galaxies in Abell~168
to check whether a case like Coma is realized 
in this recently merging cluster.
However,
we could not find an enhanced blue-galaxy distribution
between two sub-clusters.
We discuss the causes of our result,
including possibilities of surviving the hypothesis.

\end{abstract}

\section{Introduction}

Butcher \& Oemler (1978, 1984) claimed that 
the fraction of blue galaxies in rich clusters of galaxies
rapidly increases to about one fourth at a redshift of 0.5
from only a few percent at present;
this is the so-called Butcher-Oemler effect.
Subsequent spectroscopic observations
(Dressler \& Gunn 1982, 1983, 1992; Dressler {\it et al.} 1985;
Ellis {\it et al.} 1985; Lavery \& Henry 1986; Henry \& Lavery 1987;
Couch \& Sharples 1987; MacLaren {\it et al.} 1988;
Mellier {\it et al.} 1988; Fabricant {\it et al.} 1991)
have revealed that most of these blue galaxies are 
related to strong star-forming activity
(for a review, see Dressler \& Gunn 1988).
In the present-day rich clusters,
almost all cluster members are gas-poor E/S0 galaxies
and do not show star-forming activities.
Thus, there may have been a rapid evolution of cluster galaxies.
However, the mechanism responsible for this rapid evolution
is not yet clear.

Dressler \& Gunn (1983) suggested a possibility that
this rapid evolution is caused by the interaction with 
the intracluster medium (ICM);
gas-rich galaxies fall into the central region of a cluster
for the first time and they experience a rapid increase of 
ICM ``external'' pressure,
quickly exceeding the ISM ``internal'' pressure,
which triggers the starburst by compressing the molecular clouds
in galaxies (first infall model; see also Bothun \& Dressler 1986;
Dressler \& Gunn 1988; Gunn 1989).
Evrard (1991) showed from his simulations that
the first infall model can explain the Butcher-Oemler effect
semi-quantitatively.
However, it is also possible that another mechanism,
galaxy-galaxy interactions,
may play an important role in causing the Butcher-Oemler effect
(Lavery \& Henry 1988, 1994; Lavery {\it et al.} 1992).
It is necessary to examine whether the mechanism,
running into high pressure-ICM region, 
can actually induce the starbursts in cluster members.

Caldwell {\it et al.} (1993) found from
their spectroscopic observations that in the Coma cluster
strong Balmer-line-absorption galaxies reside
in the elongated region between two x-ray peaks,
the {\it main} Coma cluster and the NGC~4839 sub-cluster
(see their Fig.~17).
They pointed out that the strong Balmer-line-absorption galaxies
are similar to post-starburst galaxies 
observed in the ``Butcher-Oemler'' clusters.
Burns {\it et al.} (1994) showed from their simulations 
that the sub-cluster of Coma,
which is now seen as the second x-ray peak,
penetrated the {\it main} Coma cluster 2~Gyr ago.
They also calculated that the penetrating sub-cluster members
would spread from the center of the {\it main} cluster
in the direction of the sub-cluster,
and that their distribution is similar to that of
the Balmer-line-absorption galaxies
found by Caldwell {\it et al.} (1993).
This suggests that the sub-cluster members have traveled through
the {\it main} cluster,
and suffered a higher pressure of the ICM
inducing a burst of star formation in these galaxies.
This phenomenon can be understood in the same way 
as the first infall model;
this time passing through the {\it main} cluster gives
the same effect to galaxies as infalling toward the cluster center.
Merging clusters must be a suitable laboratory
for testing the first infall model.

We examined a distribution of blue galaxies,
regarded as star-forming galaxies,
in a merging cluster of galaxies, Abell~168.
If gas-rich galaxies have 
orbits on which a sudden change of the ICM pressure can occur,
inducing starbursts,
we would expect to see a second Coma case, i.e.,
the blue galaxies would spread in the region from
the {\it main} cluster center to the direction of the sub-cluster.
The paper is organized as follows:
the characteristics of the target cluster, 
Abell~168, are given in Sec.~2.
Observations and data reduction including the discussion
on the completeness
of our sample are given in Sec.~3.
In Sec.~4, we describe our catalog of observed galaxies.
Distribution and morphology of the ``blue'' galaxies
are described in Sec.~5 and 6, respectively.
Some comments on dominant galaxies are given in Sec.~7.
Discussion follows in Sec.~8, 
and we summarize briefly in Sec.~9.

\section{Target Cluster Abell~168}

Abell~168 has a redshift of 0.04479 and this corresponds to 131~Mpc
(we use $H_{0} = 100$~km~s$^{-1}$~Mpc$^{-1}$
and $q_{0} = 0$ throughout this paper).
This is very feasible distance
for covering a large area of the cluster by several CCD fields
and resolving the individual galaxies to be distinguished them from
stars.
The field of view of our CCD chip is about 13$'$ square,
while the order of magnitude of the cluster size, 1~Mpc,
corresponds to 27$'$.
And a seeing size is about 4$''$ (see Sec.~3) which is
fearly smaller than the typical size of small galaxies,
5~kpc $\sim$ 8$''$.
Our limiting maganitude is 19~mag(See Sec.~3). 
The distance modulus ($m-M$) of 35.7 leads to
absolute magnitude of $-$16.7~mag which is fainter than
that of SMC.
The characteristics of Abell~168 are summarized in Table~1.

Abell~168 is an example of clusters in the process of merging.
Ulmer {\it et al.} (1992) have investigated
offsets between x-ray and optical (galaxy distribution)
centers of rich clusters of galaxies.
Out of 13 clusters they studied,
Abell~168 was found to be the most extreme case;
x-ray/optical offset is 400~kpc 
(see their Fig.~2).
They suggested that Abell~168 was formed by the collision of two 
approximately equal-sized clusters with different gas-to-galaxy ratios.
Background and foreground galaxy contamination was examined
by Zabludoff {\it et al.} (1993) by measuring redshifts of galaxies
in the direction of Abell~168.
Although they found a foreground group
in the southern region of the cluster,
there is no strong contamination in the central region.
Thus, it is reliable that the x-ray/optical discrepancy is real
and x-ray multipeaks exist in Abell~168.

The hot gas is thought to be governed by the gravitational potential
of the dark matter and the typical time scale within which
the multipeaks would vanish is comparable to the sound crossing time.
Around x-ray peaks,
the temperature of the hot gas is about $3 \times 10^{7}$ K 
(David {\it et al.} 1993) and the distance between peaks
is about 400~kpc.
The calculated sound crossing time around this region is 
about 0.6~Gyr.
Therefore, not more than 0.6~Gyr could have passed after the collision
in order to see the complex x-ray distribution
and the x-ray/galaxy discrepancy;
Abell~168 is dynamically younger than Coma.
The number of member galaxies within a magnitude range 
$m_{3}$ to $m_{3}$+2 is 89 for Abell~168 and 106 for Coma,
where $m_3$ is the magnitude of the third brightest galaxy in
each cluster.
This means that Abell~168 is nearly as rich as Coma.
On the other hand,
x-ray temperature of the ICM is 2.6~keV for Abell~168
(David {\it et al.} 1993) and
this is quite lower than that for Coma, 8.3~keV.
Assuming that the mass-to-luminosity ratio of two clusters
does not so much differed from each other,
this suggests that Abell~168 is not yet virialized.
This also indicates that Abell~168 is a dynamically young system.

\section{Observations and Data Reduction}

We made $V$ and $I$ band CCD photometry of the central 0.18~deg$^{2}$
of Abell~168 in November and December 1993
using the 1.05-m Kiso Schmidt telescope (F/3.1) equipped 
with a single-chip CCD camera in the prime focus.
The CCD chip has $1000 \times 1018$ pixels and one pixel size
corresponds to 0.$''$752, giving a field-of-view
of $12.'5 \times 12.'7$.
The total observed region was covered by seven individual CCD fields;
the central position and a total exposure time 
as well as a seeing size, FWHM of stellar images in the reduced frame
of each field (about three to four arcsec),
are tabulated in Table 2.
Our search field is shown in Fig.~1
superimposed on the x-ray/galaxy map of
Fig.~2 in Ulmer {\it et al.} (1992).
The Kiso-CCD system has a good sensitivity at $V$, $R$, and $I$ bands.
The expected color change during the period of $10^{9}$ yr
after the starburst is about twice as large in ($V-I$) color
than in ($V-R$) color (see e.g., Arimoto \& Yoshii 1986).
Therefore, we chose to observe in the $V$ and $I$ bands;
filters for Johnson $V$ and Kron-Cousins $I$ band were used.

For data reduction and analysis,
we used IRAF
\footnote{IRAF is the software developed in National Optical
Astronomy Observatories.} 
and SPIRAL (Hamabe \& Ichikawa 1992) in the usual manner.
After bias-subtraction and flat-fielding
with dome-flat frames using IRAF,
sky-subtraction was made with SPIRAL.
Several object frames were combined to remove radiation events and
to improve the signal-to-noise ratio.
Object detection was made by IRAF tasks in DAOPHOT packages.
Typically about 150 objects were detected in each CCD field.
The faintest ones had magnitudes of about 20.5 and 19.5
in $V$ and $I$ bands, respectively.
We will describe the details of photometry, 
discrimination between galaxies and stars,
and completeness of the sample below.

By using photometric standard star (Feige~11) frames (Landolt 1992),
we converted CCD count rates into magnitude
($V$: Johnson, $I$: Kron-Cousins).
We made photometry by following the equations:

$V_{\rm Landolt}= v + c_{\rm V1} + c_{\rm V2}F(z) + c_{\rm V3}(V-I)$,

$I_{\rm Landolt}= i + c_{\rm I1} + c_{\rm I2}F(z) + c_{\rm I3}(V-I)$,

\noindent
where $v$ and $i$ indicate $-2.5$ log (count rate) on CCD,
$F(z)$ is the air mass function.
The coefficients $c_{\rm V1}, c_{\rm I1}, c_{\rm V2}$, and
 $c_{\rm I2}$ were determined for each night.
We used $c_{\rm V3} = 0.07$ and $c_{\rm I3} = -0.03$
for all observation nights;
these values were measured by observing five stars
with different colors in the PG~0942$-$029 field (Landolt 1992). 
We performed aperture photometry using the IRAF tasks
in APPHOT package;
the aperture radius adopted was 10~arcsec
(about 3 times point source FWHM, see Table~2)
for both $V$ and $I$ band images.
In the fourth column of Table~3
(the catalog which will be explained in Sec.~4), 
we list the size of the galaxies defined as
the major axis at 23~mag~arcsec$^{-2}$ isophote in $I$ band.
If it exceeds 20~arcsec,
we took the aperture radius of one half of the major-axis size.
We estimated the error of the photometry by comparing the data 
for a given object taken in more than two frames;
the root-mean-square of the errors in 1~mag bin
are shown in Fig.~2 (a) for $V$ band, 
and Fig.~2 (b) for $I$ band. 
A typical error of photometry is $\Delta V \sim 0.05$ mag and 
$\Delta (V-I) \sim$ 0.08~mag at $V$ = 18~mag.
Sky brightness was about 20.8 - 21.3 and
19.2 - 19.7~mag~arcsec$^{-2}$ 
in $V$ and $I$ band, respectively.

In order to distinguish galaxy images from stellar images,
we constructed two diagrams;
Diagram~A (Fig.~3) and Diagram~B (Fig.~4).
Diagram~A shows peak count versus FWHM of the images and
Diagram~B shows peak/FWHM$^{2}$ versus magnitude.
Open and filled circles in both diagrams
indicate what we could identify by eye inspection as
galaxies or foreground stars, respectively.
Crosses indicate cases,
where we could not reliably assign any category.
In both diagrams,
a galaxy sequence and a stellar sequence clearly separate
from each other and both sequences agree with the grouping
by eye inspection fairly well.
Therefore, we could distinguish galaxies from stars on CCD images
down to about 19~mag in $V$ band and about 18~mag in $I$ band.
Finally, we detected 143 galaxies brighter than $V \sim$ 19~mag.

The galaxy luminosity function in Abell~168
was studied by Dressler (1978) ($F$~band)
and Oegerle {\it et al.} (1986) ($r$~band).
Fig.~5 shows the differential luminosity function in the $V$ band
of our sample including background contamination (filled circles)
and that in the central 0.2 deg$^{2}$ by Dressler (1978)
(open squares) and by Oegerle {\it et al.} (1986) (open triangles).
Dressler (1978) and Oegerle {\it et al.} (1986)
subtracted the background-contamination by using
the field galaxy number density data in the $J$~band by Oemler (1974);
this background data reaches down to about 20~mag.
A solid line in Fig.~5 indicates
the general background count by Oemler (1974).
We converted the magnitude in different color bands assuming that 
($J-V$) = 0.3~mag, ($V-F$) = 0.7~mag, and ($V-r$) = 0.27~mag
(Oemler 1974; Oegerle {\it et al.} 1986).
Our luminosity function reproduces that by Dressler (1978) and
Oegerle {\it et al.} (1986) in the bright part ($\lesssim$ 18~mag)
and follows the background count by Oemler (1974) in the faint end
($\sim$~19~mag).
Thus, our galaxy count in the central Abell~168 region is
consistent with these previous papers;
our sample is found to be nearly complete down to about
$V \sim$ 19~mag.

Finally we show that the depth of our search is suitable.
The Butcher \& Oemler (1984) sample has a magnitude range of
about five in each cluster, and there are many blue galaxies
within this magnitude range. 
Our sample also covers the same magnitude range;
the brightest magnitude in our sample is about $V$ = 14~mag
and the faintest one is about $V$ = 19~mag.
Strong Balmer-line-absorption galaxies in Coma cluster
are found in a wide magnitude range and the faintest one
among them has a magnitude of $V$=17.2~mag,
which corresponds to $V$=18.7~mag at the distance of Abell~168.
Therefore, we searched galaxies deeply enough
to catch blue galaxies.

\section{The Catalog}

The catalog of 143 objects is given in Table~3.
An ID-number is tabulated in the first column of Table~3.
Position in 1950.0 equinox is given in the second and third columns;
the position error is about 5~arcsec.
The positions of the galaxies are determined from
those of some bright galaxies previously known;
the data are taken from NASA/IPAC Extragalactic Database (NED).
The fourth column indicates the major axis size of a galaxy
in arcsec at 23~mag~arcsec$^{-2}$ isophote in $I$ band;
1~arcsec corresponds to 610~pc.
The $V$ and $I$ magnitudes and the ($V-I$) color are given
in the fifth, sixth, and seventh column, respectively.
Eye-inspected morphology is assigned for each galaxy and
is tabulated in the eighth column.
Since most of the galaxies have sizes smaller than
four times of the point source FWHM,
we can not see the morphology in detail.
Therefore, we only assign for morphology either of
E, S0, early S (S$e$), late S (S$l$), I;
if we can see any bar structure,
we indicate this by SB$e$ or SB$l$.
If we could not assign unambiguously one of the categories,
we write for instance, S0/S$e$.
``?'' is added if the classification is uncertain.
We tabulate the distribution of morphologies in Table~4.
Here we combine barred and non-barred,
and uncertain assignments, like for instance S$e$?,
are also included into one category.
Objects assigned to two categories are divided by number
into each category;
for instance, there are five E/S0 galaxies and we count here
that there are 5/2 E's and 5/2 S0's.
From Table~4, we find that there are many spiral galaxies;
nearly two thirds are spirals and
Abell~168 is a spiral-rich cluster.
In the ninth column,
we list the ID-number given in Dressler (1980).
Radial velocities of 22 galaxies are measured
by Faber \& Dressler (1977),
Chapmann {\it et al.} (1988), and Zabludoff {\it et al.} (1993).
We list heliocentric radial velocities in the tenth column;
if more than two observations exist for one galaxy,
we take the later one.
Some comments are given in the last column.
We use many abbreviations here; see caption.

\section{Distribution of the Blue Galaxies}

Figure~6 shows the ($V-I$) color - $V$ magnitude diagram (C-M diagram).
Most galaxies have a color of 1.2 to 1.4,
and we can see a small color-magnitude (C-M) effect 
(i.e., bluer colors for fainter galaxies)
on this diagram.
Since we will examine the color distribution within a magnitude range
of over five,
the C-M effect is not negligible in defining the blue objects.
We measured the C-M effect by fitting the points on Fig.~6
having $V$ = 13.5 - 17.5~mag and ($V-I$) = 1.1 - 1.5~mag and obtained
$\Delta (V-I) / \Delta V$ = 0.032.

Figure~7 presents the color distribution of Abell~168 galaxies
after correcting the C-M effect;
we converted observed Johnson-Kron-Cousins color into
C-M effect-corrected color
as follows:

$(V-I)_{\rm corr}=(V-I) + 0.032(V-V_{\rm C})$,

\noindent
where $V_{\rm C}$ is an arbitrary value,
and here we adopt 17.0~mag as a mid point of observed $V$ magnitudes;
peak color is set to be $(V-I)_{\rm corr} = 1.2$~mag.
Since the error of color is about 0.1~mag in the worst case,
the step of color in Fig.~7 is taken to be 0.1~mag.
We can see in Fig.~7 a high concentration in 
$(V-I)_{\rm corr} =$ 1.1 to 1.3~mag, which corresponds to the typical
colors of early-type galaxies.
Objects whose colors are $(V-I)_{\rm corr} =$ 1.0 to 1.4~mag
occupy about 80~\% of the total objects. 
There is also a small blue wing in Fig.~7.
We define as a color criterion for the blue objects
that they are more than 0.2~mag bluer than the mode of
the color distribution of 1.2~mag,
i.e., bluer than $(V-I)_{\rm corr} =$ 1.0~mag.
Galaxies having blue colors of
$1.0$~mag $< (V-I)_{\rm corr} \leq$ 1.1~mag
are denoted as semi-blue.
In Fig.~6 the filled and shaded circles indicate the blue
and semi-blue objects, respectively, 
and a solid line shows the typical color of early-type galaxies,
$(V-I)_{\rm corr} =$ 1.2~mag.
In Fig.~7 the filled and shaded areas show the regions
for the blue and the semi-blue objects, respectively.
Bica {\it et al.} (1990) calculated synthetic colors
of galaxies with a partial (0.1, 1, and 10~\% mass) starburst
superimposed on old,
already existing populations.
From their model of a ten percent mass-starburst,
$\Delta (V-I)$ = 0.2~mag separates galaxies younger 
than $10^{9}$ yr after the starburst from ``old-color'' galaxies
\footnote{
Bica {\it et al.} (1990) gave the tables in Johnson system.
In converting Johnson $(V-I)$ color to Johnson-Kron-Cousins
$(V-I)$ color,
we used the relation $(V-I)_{\rm JKC}=(1/1.30)(V-I)_{\rm J}+0.01$
(Cousins 1976).
}.
Balmer-line-absorption galaxies in Coma cluster generally have blue
($B-V$) colors;
about 0.1~mag bluer in $(B-V)$ than the color mode after correcting
the C-M effect.
$\Delta (B-V) \sim$ 0.1~mag corresponds to
$\Delta (V-I) \sim$ 0.1~mag in the 10 percent-mass
partial-starburst model by Bica {\it et al.} (1990).

Figure~8 shows the distribution of galaxies in Abell~168
detected in this study
superimposed on the x-ray (0.15-4.0 keV) intensity contours
from Ulmer {\it et al.} (1992).
The blue and semi-blue objects are indicated by
the filled and shaded circles with their ID-numbers, respectively.
The size of the circles is proportional to the size of
the major axis of the galaxies,
as tabulated in the fourth column of Table~3,
though the scale is expanded twice.
Although redshift measurements by Zabludoff {\it et al.} (1993)
showed that there is no foreground contamination,
there remains a problem of some background contamination.
Since we are searching for the blue object concentration
in the cluster field,
a uniform contamination does not affect our study,
though, the contrast of the blue galaxy distribution
may become smaller.
We also list in Table~4 the blue fractions
for each morphological type.
They are 20 and 46~\%
for early spirals and late spirals, respectively.
Assuming that field galaxies equally consist of
early and late spirals,
and that the blue fraction in each morphological category
for the field galaxies is the same as that for Abell~168 members,
the blue fraction for the field galaxies would be
(20 + 46)/2 = 33~\%, almost one third. 
From Fig.~5 our sample is expected to have about
30 contamination galaxies with $V$ magnitude
fainter than about 18.5~mag.
So we have about ten blue contamination galaxies in our sample.
It contains 23 blue galaxies,
of which 14 galaxies have magnitudes
brighter than 18.5~mag and 9 galaxies are fainter than that.
Thus, all blue galaxies fainter than 18.5~mag may be
background-contamination galaxies.
The numbers with circles in Fig.~8 indicate
the objects brighter than or equal to 18.5~mag; 
these galaxies are very likely cluster members.

From Fig.~8 we can see that the blue and semi-blue 
objects spread over the cluster
and neither avoid the region between the two x-ray peaks
nor concentrate in it.
Though the number density of galaxies
near the cD galaxy (No. 46) is not large,
it seems that the blue galaxies do not tend to
distribute around the cD galaxy.
We show in Fig.~9 the blue and semi-blue galaxy fraction
as a function of distance to the cluster center
in units of the Abell radius
($R_{\rm A}$ [arcmin] $\equiv 1.7 z^{-1}$).
We assume that the cluster center is located at
the maximum of the galaxy surface number density
taken from data by Geller \& Beers (1982)
(this point is indicated by a cross in Fig.~8);
note that it is also nearly the center of our observed region.
Filled and shaded areas are for the blue and semi-blue objects,
respectively.
Fig.~9 (a) contains all galaxies and Fig.~9 (b) only
contains the brighter ($\leq$ 18.5~mag) galaxies.
It seems that the blue galaxies tend to avoid the cluster center
(within 0.1~$R_{\rm A}$)
but lie within 0.4~$R_{\rm A}$,
though the statistical error in the outer region is large.

We call the first and second x-ray peaks Point~A and Point~B,
respectively,
as shown in Fig.~8.
We assume that these two peaks represent
the centers of sub-clusters.
Though Point~B is the second x-ray peak,
it is close to the location of the maximum of
the galaxy surface number density.
The central x-ray surface brightnesses of Point~A and Point~B
are 9 $\times$ $10^{-14}$ and 3 $\times$ $10^{-14}$
erg~s$^{-1}$~cm$^{-2}$~arcmin$^{-2}$ (0.15 - 4.0~keV),
respectively (tabulated in Table~1).
Note that for Coma,
the peak x-ray surface brightness is higher, 2.3 $\times$ $10^{-13}$
erg~s$^{-1}$~cm$^{-2}$~arcmin$^{-2}$ (0.15 - 4.0~keV);
this reflects that the virialization is more developed in Coma,
as mentioned in Sec.~2.
For Coma,
the central x-ray surface brightness ratio of the main
to the sub-cluster is ten,
while for Abell~168,
the ratio is three.
Therefore, unlike the Coma case,
in Abell~168 the richness of the northern and southern sub-clusters
are not so different from each other.
A reference direction is defined by a line connecting the centers 
of both sub-clusters.
Centered on either Point~A or Point~B,
we measure angles counter-clockwise from
the reference direction mentioned above.
We performed an angular variation analysis here
with respect to both Point~A and Point~B.

Figure~10 shows the angular variation of
the blue and semi-blue galaxy fraction
in the bin of 45~degree.
Figure~10 (a) is for all galaxies in Abell~168
except for the cD galaxy at Point~A.
Filled and shaded areas are for
the blue and the semi-blue galaxies,
respectively.
Figure~10 (b) is the same as Fig.~10 (a), however,
excluding the fainter ($>$ 18.5~mag) objects.
A broken line indicates the fraction of
the blue and semi-blue galaxies;
statistically poor regions
(when the total number is less than 9)
are neglected.
We do not see an enhanced concentration of 
the blue galaxies along the line between the two x-ray peaks.
Figures~10 (c) and (d) are the same as Figs.~10 (a) and (b),
respectively,
but centered on Point~B.
Since Point~B is near the center of our observed region,
angular variation analysis is statistically better.
It is much clearer that the angular variation of
the blue galaxies is flat. 
Figure~10 (e) is for the strong Balmer line-absorption galaxies
in Coma cluster made from the data by Caldwell {\it et al.} (1993).
The center is taken to be the x-ray peak in the {\it main} cluster.
The definition of the angle is the same as that for Abell~168
described above.
We have selected galaxies in the annular region with radii
from the {\it main} cluster center from 5 to 48 arcmin;
to avoid the very crowded region we have excluded
the central region,
and to emphasize the distribution
in the region between the two x-ray peaks
we exclude the outer region.
The shaded area is for the strong Balmer-line-absorption galaxies.
We can see that the Balmer-line-absorption galaxies
frequently exist in the direction of 0\deg, that is,
to the direction of the sub-cluster.
We can not see any similar configuration of
blue galaxies in Abell~168
as that of the strong Balmer-line-absorption
galaxies in the Coma cluster.

There is another important difference between the results
presented here and by Caldwell {\it et al.} (1993).
Since the sample in Caldwell {\it et al.} (1993)
excluded late-type galaxies,
their blue galaxies in Coma are
blue early-type galaxies.
From Table~4, the number of blue early-type galaxies
in Abell~168 is small;
our blue galaxies in Abell~168 are mainly blue late-type galaxies.
Our sample contains four blue or semi-blue E/S0 galaxies
(including S0/S$e$);
No. 18, 44, 97, and 107.
They also do not tend to distribute between the two x-ray peaks.
Therefore, we found no Coma-phenomenon in Abell~168,
though we should note that our search is based only on photometry.

\section{Morphology of the Blue Galaxies}

There are nine galaxies with companions and
fifteen double galaxies
(for both of them the distance between two galaxies
are within 20 arcsec)
in our 143 galaxy sample.
``Companion'' and ``double'' are distinguished from
each other by the size of
a companion relative to a galaxy;
when the size of the companion is comparable to the galaxy,
we call it double.
We could not separate objects too close to each other in photometry;
in that case, we made photometry including the companion.
While none of nine galaxies with companions are blue or semi-blue,
eight out of fifteen (53~\%) double galaxies are blue or semi-blue.
Though seeing blurring is serious,
we found distorted morphologies in twelve galaxies
out of total 143 galaxy sample.
Eight out of the distorted galaxies (67~\%) are blue or semi-blue.
These two percentages are very high comparing
to the total blue-fraction;
(23 + 17) galaxies out of 143 (28~\%) are blue or semi-blue.
Assuming the double or distorted galaxies are 
in galaxy-galaxy (g-g) interaction,
58~\% of the g-g interacting galaxies are blue or semi-blue.
Conversely,
the blue galaxies tend to show the sign of the g-g interaction;
nearly a half of the blue or semi-blue galaxies
brighter than 18.5~mag are double or distorted galaxies.
Besides,
almost all blue or semi-blue galaxies are spirals (see Table~4).
In Abell~168, a significant fraction of the blue galaxies are
g-g interacting spirals.

\section{Comments on Some Galaxies}

In the following,
we give some comments on three dominant cluster members:

\noindent
{\bf No.39: }
This is an early-spiral and the only $IRAS$ source
in our search region.
$V$ = 15.2~mag, $I$ = 14.0~mag, and ($V-I$) = 1.2~mag.
The source name is $IRAS~F01123-0008$ and
$f_{\rm 60~\mu m}$ = 0.356~Jy.
Neither the 4.85~GHz survey by Becker {\it et al.} (1991) nor
the 1.5~GHz survey by Owen {\it et al.} (1993) detected this object.
Sopp \& Alexander (1991) showed the relation between
radio (6 cm) flux and infrared (60~$\mu$m) flux and
found two sequences;
one is for radio-quiet galaxies which are considered to be
star-forming galaxies 
($f_{\rm 6~cm}/f_{\rm 60~\mu m}$ $\sim 3 \times 10^{-3}$)
and the other is for
radio-loud galaxies 
($f_{\rm 6~cm}/f_{\rm 60~\mu m}$ $\geq$ 10).
The radio flux of galaxy No.~39 would be larger than about 3.5~Jy
if it is in the radio-loud sequence,
which should have been detected
in the radio continuum surveys mentioned above.
Therefore, galaxy No.~39 is a normal star-forming galaxy. 

Assuming that the far-infrared emission of galaxy No.~39
originates from the star-forming activity,
we can calculate the present star formation rate.
Kennicutt {\it et al.} (1994) derived the relation between
the present star formation rate (SFR) and the \HA luminosity:
SFR~[$M_{\odot}$ yr$^{-1}$]/$L$(\HA )~[$10^{41}$~erg~s$^{-1}$] = 1.36.
Young {\it et al.} (1989) found
the relation between the infrared luminosity
and the \HA luminosity:
$L_{\rm IR}$(1 - 500 $\mu$m) = $3.0 \times 10^{2}~L$(\HA ).
These relations and 
$L_{\rm IR}$(1 - 500 $\mu$m)/$L_{\rm FIR}$(42.5 - 122.5 $\mu$m) = 1.5
for the (60 - 100~$\mu$m) color of this object
({\it Cataloged Galaxies and Quasars in the IRAS Survey} 1985)
lead to a present 
SFR = 1.9~$M_{\odot}$ yr$^{-1}$ for this object. 
The present SFRs in other members are, therefore, 
lower than about 2~$M_{\odot}$ yr$^{-1}$.

\noindent
{\bf No.46: } 
This is a cD galaxy, the brightest member;
$V$ = 13.9~mag, $I$ = 12.6~mag, and ($V-I$) = 1.3~mag.
The position of this galaxy corresponds to the x-ray peak,
the Point~A.
This galaxy is not detected in $IRAS$ survey nor 
4.85 and 1.5~GHz radio continuum surveys mentioned above.

\noindent
{\bf No.81: }
This is an elliptical and the second brightest member;
$V$ = 15.0~mag, $I$ = 13.7~mag, and ($V-I$) = 1.3~mag.
Its position is near to the second x-ray peak,
the Point~B.
This is the only radio source detected so far
in our searched region;
$S$(4.85~GHz) = 42~mJy (Becker {\it et al.} 1991) and
$S$(1.5~GHz) = 49~mJy (Owen {\it et al.} 1993).
The spectral index of $\alpha$ ($F_{\nu} \propto \nu^{\alpha}$)
is $-0.13$, 
which indicates a fairly flat spectrum.
Assuming ($B-V$) = 1.0~mag,
$R$ $\equiv$ $S$(6 cm)/$S$(4400 \AA ) = 23
(the definition of $R$ is from Kellermann {\it et al.} 1989)
this galaxy is in the mid point between
radio-loud and radio-quiet objects.
If the radio flux of this galaxy is
from star-forming activity only,
$f_{\rm 60~\mu m}$ would be about 10~Jy and
this is bright enough to be detected in the $IRAS$ survey.
If the radio flux is from some other powerful activities,
$f_{\rm 60~\mu m}$ would be fainter than several mJy 
and thus be too faint to detect in the $IRAS$ survey.
Probably this galaxy belongs to the radio-loud objects.

Many of the spirals in Abell~168 are located near or north
of the second brightest member, No.~81, Point~B. 
This distribution is similar to that of sub-cluster members
after merging in the simulation of Burns {\it et al.} (1994);
it may be an indirect evidence of the merger event.

\section{Discussion}

In Abell~168 we did not find a concentration
of blue galaxies at any place,
particularly not in the region between the two x-ray peaks.
Though our search is based on photometry only,
we could not find the Coma-phenomenon in Abell~168.
There are some possibilities to explain our results as follows:

\noindent
1. almost all infalling galaxies were gas-poor, 
therefore, they did not have the ability to burst,

\noindent
2. the orbits on which the ICM pressure increases are not realized,

\noindent
3. we did not pick up all post-starburst 
(strong Balmer-line-absorption) galaxies, or

\noindent
4. star-forming activity is not induced by running into 
the dense ICM regions.

The hypothesis suggested by Dressler \& Gunn (1983) needs both
the gas-rich galaxies and the running orbits
toward the dense ICM regions.
If almost all galaxies had been gas-poor,
blue galaxies could not appear.
There is no \HI or CO data for Abell~168 members so far,
therefore,
we do not know whether these galaxies are really gas-poor,
or whether they are gas-rich galaxies which do not show 
strong star-forming activity.
Neutral or molecular gas observations are necessary
to check this.
However, Abell~168 is a spiral-rich cluster and
it is improbable that almost all members are gas-poor.

The second possibility is that 
the orbits may not be realized in Abell~168
because of a lack of the effective ICM compression.
The x-ray contrast of two sub-clusters in Abell~168
is about one third of that in Coma,
as mentioned in Sec.~5.
This may cause much calmer environmental change 
for merging members in Abell~168 than for those in Coma.
To clarify the environmental history of member galaxies,
high-quality x-ray hot gas observations are necessary.

The strong Balmer-line-absorption galaxies
in the Coma cluster studied by Caldwell {\it et al.} (1993) 
are considered to be in the post-starburst phase,
however, some of them have already entered into
a evolved stage as ``weak'' Balmer-line-absorption systems.
Though most of the blue galaxies 
(bluer than $\Delta (B-V) \sim$ 0.1~mag
after correcting the C-M effect)
in Coma are the post-starburst galaxies,
about one third of all the post-starburst galaxies are not that blue.
We may miss some of the post-starburst galaxies in Abell~168.
Dressler \& Gunn (1992) investigated the color distribution
of emission galaxies, post-starburst galaxies, and passive galaxies
in seven rich clusters.
From their Fig.~22, generally the post-starburst galaxies
(strong Balmer-line-absorption galaxies)
have blue colors,
however, some of them are as red as passive, typical E/S0 galaxies.
In order to pick up the complete post-starburst galaxy population,
it is necessary to make spectroscopic observations.

The final possibility might be the case {\it in Abell~168}.
Our results probably show that the rapid increase of ICM pressure
for the galaxies does not 
work for starburst trigger {\it in Abell~168}.
However, here we observed only one cluster and
we have to check in many other merging clusters.
At present we can not say whether
our results show that galaxy-ICM interaction does not affect
starburst triggering or
whether they are due to other causes.

\section{Summary}

We examined whether the first infall model is valid for 
triggering starbursts in cluster members or not;
the model is only satisfied 
when all of the following three conditions exist:

\noindent
(1) gas-rich galaxies,

\noindent
(2) running orbit into the dense ICM regions resulting in the rapid
increase of the ``external'' pressure for the galaxies, and

\noindent
(3) starburst triggering by compression of molecular clouds by
a raised ``external'' pressure.

The coma-phenomenon,
post-starburst galaxies spread between two x-ray peaks,
can be interpreted as the result of a cluster merger.
Some galaxies would be rushing into the dense ICM regions 
in the merging cluster.
We analyzed the distribution of the blue galaxies,
regarded as the star-forming galaxies, 
in the young merging cluster of galaxies, Abell~168
to see if there is a second example of the Coma-phenomenon.
However, we could not find a blue galaxy distribution in Abell~168 
to support the model mentioned above.
Probably the mechanism does not work effectively as a
starburst trigger (negation of (3)),
however, there remain some other possibilities.
For instance,
galaxies may have been already gas-free (contradicting (1)),
the orbits on which the ICM pressure increases are not realized
(contradicting (2)), or
we have missed some post-starburst galaxies
which are not blue,
and not identified as such in our search.
We need cold and hot X-ray gas observations as well as spectroscopic
observations,
and much data from many other merging clusters for further study.
Some starburst galaxies may have left evidences
for the galaxy-ICM interaction
like metal-enriched ``islands'' in the hot gas around the galaxies.
High spatial-resolution X-ray observations could be another
effective way to search for burst-induced galaxies.

We also found in Abell~168  that nearly half of the blue and semi-blue
galaxies are spirals in galaxy-galaxy interaction.

\acknowledgments

We would like to thank Mamoru Sait\={o}, Rainer Spurzem,
Takashi Ichikawa,
Shin Mineshige, and Chiharu Ishizaka for valuable discussions,
as well as the Kiso Observatory staff members for their help in 
the observations. 
We also thank to Masaru Hamabe and Nobunari Kashikawa
for giving us the latest version of SPIRAL and
showing some useful techniques on distinguishing between
galaxy and stellar images,
respectively.
This research has made use of
the NASA/IPAC Extragalactic Database (NED)
which is operated by the Jet Propulsion Laboratory, Caltech,
under contact with the National Aeronautics and Space Administration.


\clearpage

\begin{figure}

\centerline{FIGURE CAPTIONS}

\caption{
Our search region in Abell~168,
superimposed on Fig.~2 in Ulmer {\it et al.} (1992).
Open circles indicate bright galaxies identified by Dressler (1980);
their ID-numbers are given in the ninth column of Table~3,
and contours are for x-ray brightness isophotes observed by
$Einstein~IPC$ (0.15 - 4.0 keV).
A filled square and a filled triangle denote the x-ray peak and
the center of galaxy distribution by Ulmer {\it et al.} (1992),
respectively.
}

\caption{
The error of photometry 
for (a) $V$ band data and (b) $I$ band data
estimated by comparing data
for a given object taken in more than two frames.
We show the root-mean-square of measurements
for the errors in 1~mag bin.
Most of the measurements have an error of less than 0.05~mag.
}

\caption{
An example of Diagram~A,
peak count of the image versus FWHM of the image size diagram.
This diagram is for the field No. 7 in $V$ band.
Open and filled circles show images which are identified
by eye-inspection as galaxies and stars, respectively.
Crosses indicate the images which are difficult to be discriminated
between galaxies and stars.
}

\caption{
The same as Fig.~3
but for (peak count)/FWHM$^{2}$ versus magnitude diagram 
(Diagram~B).
}

\caption{
The differential luminosity function of Abell~168 galaxies.
Numbers are normalized to what is expected to be in 0.2 deg$^{2}$.
Filled circles represent our sample. 
Note that our sample contains the background contamination.
Open squares and open triangles show
Dressler (1978) and Oegerle {\it et al.} (1986) sample, respectively.
A solid line shows field galaxy count by Oemler (1974),
which was used for the background-contamination correction in
Dressler (1978) and Oegerle {\it et al.} (1986).
}

\caption{
($V-I$) color - $V$ magnitude diagram for Abell~168 galaxies
(C-M diagram).
A solid line shows the typical color for early-type galaxies,
the C-M effect.
The blue and semi-blue objects are shown by the filled and shaded 
circles, respectively.
}

\end{figure}

\clearpage

\begin{figure}

\caption{
($V-I$)$_{\rm corr}$ color distribution after correcting the
C-M effect.
Filled and shaded regions correspond to the blue and semi-blue objects,
respectively.
We choose the color of $(V-I)_{\rm corr}$ = 1.2~mag
to be peak of the color,
shown by a dashed line.
Note that this color corresponds to the solid line in Fig.~5.
}

\caption{
Distribution of all galaxies and blue galaxies
detected in our survey.
The blue and semi-blue galaxies are shown 
by the filled and shaded circles, respectively.
The ID-numbers are given for the blue and semi-blue objects
and some other interesting objects mentioned in text.
The numbers with circles are for the objects brighter than 18.5~mag.
The size of circles is proportional to the size of major axis,
tabulated in the third column of Table~3.
Note that the size scale is expanded twice compared to
the galaxy separation scale for comprehension.
Dashed curves indicate the x-ray (0.15-4.0 keV)
contours by Ulmer {\it et al.} (1992).
A cross indicates the cluster center used for Fig.~10,
the maximum of galaxy surface number density 
by Geller \& Beers (1982).
}

\caption{
The fraction of blue and semi-blue galaxies
with the cluster-centric radius
for (a) all 143 galaxies and for (b) only the brighter ($\leq$ 18.5~mag)
109 galaxies.
The center is taken to be the maximum
of the galaxy surface number density
by Geller \& Beers (1982), 
shown by the cross in Fig.~7;
this point is located near the center of our search region.
A broken line indicates the fraction of
the blue and semi-blue galaxies;
the scale is shown on the right side.
Note that we searched only in the central region of the cluster.
The Abell radius ($R_{\rm A}$ [arcmin] = 1.7~$z^{-1}$)
of Abell~168 is 38~arcmin.
The blue and semi-blue galaxies tend to avoid the central region
with a radius of 0.1~$R_{\rm A}$.
Though the statistical error is large,
the blue and semi-blue galaxies seem to
concentrate within 0.4~$R_{\rm A}$.
}

\end{figure}

\clearpage

\begin{figure}

\caption{
The angular variation of the fraction of the blue galaxies.
(a) 
For the blue and semi-blue galaxies in Abell~168 from our data.
Center is taken to be Point~A.
Filled area is for the blue galaxies and
shaded area is for the semi-blue galaxies.
A broken line indicates the fraction of
the blue and semi-blue galaxies;
the scale is shown on the right side.
If the total number in a cone region is less than 9,
we omit calculation of the fraction in that division
because of poor statistics.
No significant over-distribution of the blue galaxies is seen
in the angle of 0\deg .
(b) 
The same as (a), however, excluding the fainter ($>$ 18.5~mag) objects.
Again we can not see the over-distribution of
the blue and semi-blue galaxies
in the angle of 0\deg .
(c)(d) 
The same as (a)(b), 
however, centered on Point~B.
The flatness of the angular variation of
the blue galaxies is seen much clearly.
(e) 
For the strong Balmer-line-absorption galaxies in Coma cluster;
the data are taken from Caldwell {\it et al.} (1993).
Shaded area is for the strong Balmer-line-absorption galaxies.
Enhanced distribution in the angle of 0\deg , 
the direction to the sub-cluster is seen clearly.
}

\end{figure}

\end{document}


\newcommand{\HI}{H~{\small I} }
\newcommand{\HII}{H~{\small II} }
\newcommand{\HH}{H$_{2}$ }
\newcommand{\HA}{H$\alpha$ }

\centerline{TABLE CAPTIONS}

\bigskip

T{\small ABLE}~1. Characteristics of Abell~168.

T{\small ABLE}~2. A journal of observations.

T{\small ABLE}~3. Catalog of galaxies in Abell 168.

T{\small ABLE}~4. Distribution of the morphology.

\clearpage

\begin{table*}
\begin{center}

T{\small ABLE}~1. Characteristics of Abell~168.

\bigskip

\begin{tabular}{lllc}
 \tableline
 \tableline
 Center & $\alpha$(1950.0)               & $01^{\rm h}12^{\rm m}.6$ & (1) \\
        & $\delta$(1950.0)               & $-$00\deg 01$'$           & (1) \\
        & $\alpha$(2000.0)               & $01^{\rm h}15^{\rm m}.2$ & (1) \\
        & $\delta$(2000.0)               & 00\deg 14$'$              & (1) \\
        & $l$                            & 135.\deg 67               & (1) \\
        & $b$                            & $-$62.\deg 04             & (1) \\
 Type   & Bautz-Morgan                   & II-III:                  & (1) \\
        & Rood-Sastry                    & I$_{\rm C}$              & (2) \\
 \multicolumn{2}{l}{Count\tablenotemark{a}}
                                         & 89                       & (1) \\
 \multicolumn{2}{l}{Radial velocity}     & 13428 km s$^{-1}$        & (3) \\
 \multicolumn{2}{l}{Richness class}      & 2                        & (1) \\
 \multicolumn{2}{l}{Distance class}      & 3                        & (1) \\
 \multicolumn{2}{l}{$m_{10}$\tablenotemark{b}}
                                         & 15.4                     & (1) \\
 \multicolumn{2}{l}{Radial velocity dispersion} & 461 km s$^{-1}$   & (3) \\
 \multicolumn{2}{l}{Diameter of the cluster} & 3 Mpc                & (5) \\
 \multicolumn{2}{l}{Mass of the cluster\tablenotemark{c}}
 & 3.5 $\times 10^{14} M_{\odot}$  &     \\
 \multicolumn{2}{l}{Temperature of the hot gas}
 & 3.0 $\times 10^{7}$ K (2.6 keV) & (4) \\
 \multicolumn{2}{l}{Peak X-ray surface brightness\tablenotemark{d}} & & \\
  & Point A & 9 $\times$ $10^{-14}$ erg s$^{-1}$ cm$^{-2}$ arcmin$^{-2}$ & \\
  & Point B & 3 $\times$ $10^{-14}$ erg s$^{-1}$ cm$^{-2}$ arcmin$^{-2}$ & \\
 \tableline
 \tableline
\end{tabular}

\bigskip

Notes to T{\small ABLE}~1

\end{center}

\rightskip=1.5cm
\leftskip=1.5cm

$^{\rm a}$
Background-corrected count of cluster members in the magnitude range
$m_{3}$ to $m_{3}$+2,
where $m_{2}$ is the magnitude of the third brightest galaxy.

$^{\rm b}$
The red magnitude of the tenth brightest cluster member.

$^{\rm c}$
The cluster mass $M$ is derived from the equation,
$M = 3 R \sigma_{\rm gal}^{2} / G$.
$R$ is the diameter of the cluster and
$\sigma_{\rm gal}$ is the radial velocity dispersion of the galaxies;
both values are taken from those tabulated here.

$^{\rm d}$
Measured by $Einstein~IPC$ (0.15 - 4.0~keV).
We referred to {\it Einstein Observatory Catalog of IPC X-ray
Sources Volume 1} (1990).

References

(1) Abell {\it et al.} (1989)

(2) Struble \& Rood (1982)

(3) Zabludoff {\it et al.} (1993)

(4) David {\it et al.} (1993)

(5) Zwicky, Karpowicz, \& Kowal (1965)

\end{table*}

\rightskip=0cm
\leftskip=0cm

\clearpage

\begin{table*}
\begin{center}

T{\small ABLE}~2. A journal of observations.

\bigskip

\begin{tabular}{ccccc}
 \tableline
 \tableline
No.& Field center           & Total exposure& Observation& Seeing size    \\
   & $\alpha (1950.0)\delta$& time (sec)    & Date       &
 (arcsec)\tablenotemark{a} \\
 \tableline
   & \ $^{\rm h}$ \ \ $^{\rm m}$ \ \ $^{\rm s}$ \ \ \ \deg \ \ $'$
                                & ($V,I$)      & (1993)      & ($V,I$)  \\
 1 & 01\ 12\ 29,\ \ \ 0\ 12.\ 9 & \ 300, \ 300 & 15 November & 3.7, 3.2 \\
 2 & 01\ 12\ 28,\ \ \ 0\ 05.\ 6 & \ 300, \ 300 & 15 November & 3.8, 3.5 \\
 3 & 01\ 13\ 01,\ \ \ 0\ 05.\ 3 & 1500, 1500   & 15 November & 3.5, 4.1 \\
 4 & 01\ 13\ 01,\  $-$0\ 02.\ 3 & 2700, 3300   & 08 December & 3.1, 3.0 \\
 5 & 01\ 11\ 58,\ \ \ 0\ 12.\ 9 & 2100, 2100   & 09 December & 4.2, 4.2 \\
 6 & 01\ 13\ 02,\  $-$0\ 06.\ 6 & 1200, 1200   & 12 December & 2.9, 2.9 \\
 7 & 01\ 12\ 28,\  $-$0\ 02.\ 3 & 2700, 2100   & 12 December & 3.1, 2.9 \\
 \tableline
 \tableline
\end{tabular}

\bigskip

Notes to T{\small ABLE}~2

\end{center}

\rightskip=2cm
\leftskip=2cm

$^{\rm a}$
FWHM of stellar images in the reduced frames.

\end{table*}

\rightskip=0cm
\leftskip=0cm

\clearpage

\begin{table*}
\begin{center}

T{\small ABLE}~3. Catalog of galaxies in Abell 168.

\bigskip

\begin{tabular}{rrrcccccccl}
 \tableline
 \tableline
 No. & \multicolumn{2}{c}{Position}                 & Size & $V$ & $I$ &($V-I$)
 & Morp.  & D80          & Radial   & Notes \\
     & \multicolumn{2}{c}{$\alpha (1950.0) \delta$} &      &     &     &      
 &        & No.          & velocity &       \\
 (1) & (2) \ \ \                            & (3) \ \        &
 (4)    & (5) & (6) & (7) & (8) & (9) & (10)          & \ \ (11) \\ 
 \tableline
     & $^{\rm h}$ \ $^{\rm m}$ \ $^{\rm s}$ & \deg \ \ $'$ \ &
 [$''$] &     &     &     &     &     & [km s$^{-1}$] &          \\
  1 & 1\ 11\ 39 &    0\ 17.1 &  5 & 18.8 & 17.8 & 1.0 &S$e$  &     &       & 
      sb \\
  2 & 1\ 11\ 42 &    0\ 07.2 & 11 & 17.5 & 16.3 & 1.2 &E     &     &       &
      COM \\
  3 & 1\ 11\ 47 &    0\ 12.9 & 19 & 16.9 & 15.8 & 1.2 &E/S0  &  94 &       & \\
  4 & 1\ 11\ 48 &    0\ 08.7 &  6 & 18.8 & 17.5 & 1.3 &S$e$  &     &       & \\
  5 & 1\ 11\ 48 &    0\ 09.3 & 21 & 16.3 & 15.1 & 1.2 &E     &  84 &       & \\
  6 & 1\ 11\ 49 &    0\ 16.7 & 33 & 15.2 & 14.2 & 0.9 &S$e$  & 102 & 12486 &
      b, b-disk \\
  7 & 1\ 11\ 53 &    0\ 12.8 & 16 & 16.8 & 15.6 & 1.3 &E     &  93 &       & \\
  8 & 1\ 11\ 53 &    0\ 15.5 & 10 & 17.8 & 16.7 & 1.1 &S$e$? &     &       &
      COM ? \\
  9 & 1\ 11\ 54 &    0\ 16.2 & 17 & 16.7 & 15.6 & 1.2 &E/S0  & 101 &       & \\
 10 & 1\ 11\ 55 &    0\ 12.8 & 11 & 17.7 & 16.4 & 1.3 &E     &     &       & \\
 11 & 1\ 11\ 55 &    0\ 13.1 &  8 & 19.1 & 17.5 & 1.6 &S$e$  &     &       & \\
 12 & 1\ 11\ 56 &    0\ 12.7 & 10 & 18.2 & 17.0 & 1.2 &E     &     &       &
      DBL \\
 13 & 1\ 11\ 56 &    0\ 16.3 & 21 & 16.3 & 15.1 & 1.2 &S0    & 100 &       & \\
 14 & 1\ 12\ 00 &    0\ 12.9 &  9 & 17.7 & 16.6 & 1.1 &E     &     &       & \\
 15 & 1\ 12\ 04 &    0\ 01.1 &  5 & 18.7 & 17.8 & 0.9 &S$e$/S$l$&  &       &
      b \\
 16 & 1\ 12\ 04 &    0\ 10.6 &  3 & 19.0 & 18.4 & 0.6 &S$e$/I&     &       &
      LSB, DBL(17), b \\
 17 & 1\ 12\ 05 &    0\ 10.6 &  8 & 18.5 & 17.6 & 0.9 &S$e$  &     &       &
      DBL(16), b \\
 18 & 1\ 12\ 05 &    0\ 12.5 & 23 & 16.1 & 15.0 & 1.1 &S0    &  92 & 13360 &
      sb \\
 19 & 1\ 12\ 08 & $-$0\ 07.6 & 17 & 17.0 & 15.8 & 1.2 &S0    &  42 &       &
      barred ? \\
 20 & 1\ 12\ 08 & $-$0\ 00.6 & 13 & 18.0 & 16.9 & 1.1 &S$e$  &     &       & \\
 21 & 1\ 12\ 09 &    0\ 01.9 & 29 & 15.8 & 14.7 & 1.1 &S$e$  &  72 &       &
      dist, sb \\
 22 & 1\ 12\ 10 &    0\ 17.3 &  7 & 18.9 & 17.7 & 1.2 &S$e$/I&     &       & \\
 23 & 1\ 12\ 11 &    0\ 16.3 &  8 & 18.2 & 17.1 & 1.1 &S$e$? &     &       & \\
 24 & 1\ 12\ 12 &    0\ 09.8 &  7 & 19.0 & 17.7 & 1.3 &?     &     &       &
      red \\
 25 & 1\ 12\ 12 &    0\ 13.4 &  7 & 16.0 & 14.7 & 1.4 &?     &     &       &
      DBL \\
 26 & 1\ 12\ 13 &    0\ 15.6 & 34 & 15.4 & 14.1 & 1.3 &S$e$  &  99 & 13597 &
      r-circum-nucleus \\
 27 & 1\ 12\ 13 &    0\ 18.7 & 11 & 18.1 & 16.7 & 1.4 &SB$e$ &     &       &
      r-bar \\
 28 & 1\ 12\ 14 & $-$0\ 01.7 & 22 & 16.1 & 14.9 & 1.2 &S0    &  71 & 13975 &
      COM ? \\
 29 & 1\ 12\ 14 &    0\ 04.0 &  9 & 18.4 & 17.6 & 0.8 &S$e$  &     &       &
      dist, b \\
 30 & 1\ 12\ 17 &    0\ 17.1 &  5 & 17.0 & 15.9 & 1.2 &S$e$  &  98 &       & \\
 \tableline
\end{tabular}
\end{center}
\end{table*}

\clearpage

\begin{table*}
\begin{center}

T{\small ABLE}~3. (continued)

\bigskip

\begin{tabular}{rrrcccccccl}
 \tableline
 No. & \multicolumn{2}{c}{Position}                 & Size & $V$ & $I$ &($V-I$)
 & Morp.  & D80          & Radial   & Notes \\
     & \multicolumn{2}{c}{$\alpha (1950.0) \delta$} &      &     &     &      
 &        & No.          & velocity &       \\
 (1) & (2) \ \ \                            & (3) \ \        &
 (4)    & (5) & (6) & (7) & (8) & (9) & (10)          & \ \ (11) \\ 
 \tableline
     & $^{\rm h}$ \ $^{\rm m}$ \ $^{\rm s}$ & \deg \ \ $'$ \ &
 [$''$] &     &     &     &     &     & [km s$^{-1}$] &          \\
 31 & 1\ 12\ 19 &    0\ 00.6 &  7 & 18.4 & 17.1 & 1.2 &S$e$  &     &       & \\
 32 & 1\ 12\ 19 &    0\ 02.8 & 28 & 15.2 & 13.9 & 1.3 &E     &  70 & 13294 & \\
 33 & 1\ 12\ 19 &    0\ 07.8 &  7 & 18.6 & 17.5 & 1.1 &S$e$  &     &       & \\
 34 & 1\ 12\ 20 &    0\ 14.5 & 25 & 15.7 & 14.5 & 1.3 &S$e$  &  91 &       &
      r-bar ? \\
 35 & 1\ 12\ 21 &    0\ 02.2 & 12 & 17.4 & 16.2 & 1.1 &S0/S$e$& 69 &       & \\
 36 & 1\ 12\ 21 &    0\ 12.7 & 13 & 17.5 & 16.3 & 1.2 &E/S0  &     &       &
      dist \\
 37 & 1\ 12\ 22 &    0\ 02.5 &  8 & 19.0 & 17.6 & 1.4 &S$e$  &     &       & \\
 38 & 1\ 12\ 22 &    0\ 07.8 & 15 & 16.9 & 15.6 & 1.2 &E     &  83 &       & \\
 39 & 1\ 12\ 23 & $-$0\ 08.0 & 31 & 15.2 & 14.0 & 1.2 &S$e$  &  41 & 12485 &
      r-bar \\
 40 & 1\ 12\ 23 & $-$0\ 03.4 &  8 & 18.7 & 18.2 & 0.6 &I?    &     &       &
      DBL(47), b \\
 41 & 1\ 12\ 23 &    0\ 10.6 & 13 & 17.8 & 16.4 & 1.4 &E     &     &       & \\
 42 & 1\ 12\ 23 &    0\ 16.0 & 13 & 17.4 & 16.2 & 1.2 &E     &     &       & \\
 43 & 1\ 12\ 24 & $-$0\ 06.9 &  8 & 18.5 & 17.2 & 1.3 &S$e$  &     &       & \\
 44 & 1\ 12\ 24 & $-$0\ 06.1 &  8 & 18.6 & 17.6 & 1.0 &S0/S$e$&    &       &
      sb \\
 45 & 1\ 12\ 24 &    0\ 00.3 & 11 & 17.5 & 16.3 & 1.2 &E     &     &       & \\
 46 & 1\ 12\ 24 &    0\ 10.0 & 55 & 13.9 & 12.6 & 1.3 &E (cD)&  90 & 13439 & \\
 47 & 1\ 12\ 25 & $-$0\ 03.3 &  8 & 17.9 & 17.2 & 0.7 &S$e$  &     &       &
      DBL(40), b \\
 48 & 1\ 12\ 25 & $-$0\ 00.8 &  8 & 18.5 & 17.5 & 1.0 &S$e$? &     &       &
      sb \\
 49 & 1\ 12\ 25 &    0\ 11.6 &  8 & 18.7 & 17.4 & 1.3 &S$e$  &     &       & \\
 50 & 1\ 12\ 26 & $-$0\ 07.3 & 14 & 17.6 & 16.3 & 1.4 &S$e$  &     &       & \\
 51 & 1\ 12\ 26 & $-$0\ 02.8 & 11 & 18.0 & 16.9 & 1.1 &S$e$  &     &       & \\
 52 & 1\ 12\ 26 & $-$0\ 00.2 & 28 & 15.4 & 14.1 & 1.3 &SB$e$ &  68 & 14540 &
      r-bar \\
 53 & 1\ 12\ 26 &    0\ 06.5 & 10 & 18.5 & 17.2 & 1.3 &?     &     &       & \\
 54 & 1\ 12\ 26 &    0\ 15.3 &  7 & 18.9 & 17.8 & 1.0 &?     &     &       & \\
 55 & 1\ 12\ 28 &    0\ 00.2 & 23 & 16.4 & 15.2 & 1.2 &S$e$  &  66 &       & \\
 56 & 1\ 12\ 28 &    0\ 02.9 & 21 & 16.7 & 15.5 & 1.2 &S$e$  &  67 &       & \\
 57 & 1\ 12\ 30 &    0\ 08.4 & 32 & 15.0 & 13.8 & 1.2 &S$e$  &  82 & 13340 & \\
 58 & 1\ 12\ 30 &    0\ 17.3 &  8 & 19.0 & 17.7 & 1.3 &I?    &     &       & \\
 59 & 1\ 12\ 31 &    0\ 03.7 & 31 & 15.7 & 14.4 & 1.3 &S$e$  &  81 & 12625 & \\
 60 & 1\ 12\ 32 & $-$0\ 00.9 & 10 & 18.5 & 17.3 & 1.2 &S$e$  &     &       & \\
 \tableline
\end{tabular}
\end{center}
\end{table*}

\clearpage

\begin{table*}
\begin{center}

T{\small ABLE}~3. (continued)

\bigskip

\begin{tabular}{rrrcccccccl}
 \tableline
 No. & \multicolumn{2}{c}{Position}                 & Size & $V$ & $I$ &($V-I$)
 & Morp.  & D80          & Radial   & Notes \\
     & \multicolumn{2}{c}{$\alpha (1950.0) \delta$} &      &     &     &      
 &        & No.          & velocity &       \\
 (1) & (2) \ \ \                            & (3) \ \        &
 (4)    & (5) & (6) & (7) & (8) & (9) & (10)          & \ \ (11) \\ 
 \tableline
     & $^{\rm h}$ \ $^{\rm m}$ \ $^{\rm s}$ & \deg \ \ $'$ \ &
 [$''$] &     &     &     &     &     & [km s$^{-1}$] &          \\
 61 & 1\ 12\ 33 &    0\ 09.8 &  7 & 18.3 & 17.5 & 0.7 &?     &     &       &
      b \\
 62 & 1\ 12\ 34 & $-$0\ 02.8 &  9 & 18.7 & 17.8 & 0.9 &S$e$  &     &       &
      b \\
 63 & 1\ 12\ 34 & $-$0\ 02.1 & 23 & 16.4 & 15.3 & 1.1 &S$e$  &  63 &       &
      dist, sb \\
 64 & 1\ 12\ 34 &    0\ 11.9 & 27 & 15.6 & 14.3 & 1.3 &S0/S$e$& 89 & 12560 & \\
 65 & 1\ 12\ 35 &    0\ 00.3 & 22 & 16.0 & 14.8 & 1.2 &E     &  62 &       & \\
 66 & 1\ 12\ 35 &    0\ 02.1 & 21 & 16.5 & 15.5 & 1.1 &S$e$  &  64 & 13658 &
      DBL(67), sb \\
 67 & 1\ 12\ 35 &    0\ 02.6 & 21 & 16.4 & 15.2 & 1.2 &S$e$  &  65 & 14027 &
      DBL(66), swb \\
 68 & 1\ 12\ 36 &    0\ 18.2 & 26 & 15.4 & 14.1 & 1.3 &S$e$  &  97 &       & \\
 69 & 1\ 12\ 37 &    0\ 02.8 &  9 & 18.6 & 17.6 & 1.0 &S$e$  &     &       &
      sb \\
 70 & 1\ 12\ 38 & $-$0\ 06.6 &  7 & 18.6 & 17.5 & 1.1 &?     &     &       & \\
 71 & 1\ 12\ 39 & $-$0\ 12.7 & 11 & 18.4 & 17.3 & 1.2 &E     &     &       & \\
 72 & 1\ 12\ 39 &    0\ 03.4 &  7 & 18.9 & 17.4 & 1.5 &S$e$  &     &       & \\
 73 & 1\ 12\ 39 &    0\ 18.2 &  9 & 17.8 & 16.7 & 1.1 &?     &     &       & \\
 74 & 1\ 12\ 40 &    0\ 01.6 & 19 & 16.9 & 15.8 & 1.1 &S0    &  61 &       & \\
 75 & 1\ 12\ 40 &    0\ 04.6 & 21 & 15.9 & 14.6 & 1.3 &E     &  80 & 13453 & \\
 76 & 1\ 12\ 41 & $-$0\ 01.4 &  9 & 17.8 & 16.6 & 1.3 &E     &     &       & \\
 77 & 1\ 12\ 41 &    0\ 04.4 & 20 & 16.4 & 15.3 & 1.1 &S$e$/S$l$&79&       & \\
 78 & 1\ 12\ 41 &    0\ 13.2 & 12 & 17.6 & 16.7 & 0.9 &S$e$  &  95 &       &
      b \\
 79 & 1\ 12\ 42 & $-$0\ 09.1 &  8 & 18.7 & 17.5 & 1.1 &S$e$? &     &       & \\
 80 & 1\ 12\ 42 & $-$0\ 08.6 & 14 & 18.4 & 17.2 & 1.2 &S$e$  &     &       &
      DBL\\
 81 & 1\ 12\ 42 & $-$0\ 03.1 & 35 & 15.0 & 13.7 & 1.3 &E     &  40 & 13553 & \\
 82 & 1\ 12\ 43 & $-$0\ 10.7 & 22 & 16.4 & 15.3 & 1.1 &S$e$  &  28 &       &
      dist, sb \\
 83 & 1\ 12\ 43 & $-$0\ 06.2 & 18 & 17.1 & 16.1 & 1.0 &S$e$  &     &       &
      dist, b \\
 84 & 1\ 12\ 43 & $-$0\ 04.8 & 32 & 15.1 & 13.8 & 1.3 &E     &38,39& 13420 &
      DBL \\
 85 & 1\ 12\ 43 & $-$0\ 02.0 &  8 & 18.6 & 17.3 & 1.2 &?     &     &       & \\
 86 & 1\ 12\ 43 & $-$0\ 01.7 & 17 & 17.3 & 16.1 & 1.2 &S$e$  &  60 &       & \\
 87 & 1\ 12\ 43 &    0\ 00.8 &  7 & 18.6 & 17.8 & 0.8 &S$l$/SB$l$& &       &
      LSB, b \\
 88 & 1\ 12\ 43 &    0\ 02.8 & 13 & 17.9 & 17.0 & 0.8 &S$e$? &     &       &
      b \\
 89 & 1\ 12\ 44 & $-$0\ 10.0 & 15 & 18.4 & 17.6 & 0.8 &S$l$  &     &       &
      LSB, b \\
 90 & 1\ 12\ 44 & $-$0\ 06.0 & 15 & 16.6 & 15.3 & 1.3 &E     &  37 &       & \\
 \tableline
\end{tabular}
\end{center}
\end{table*}

\clearpage

\begin{table*}
\begin{center}

T{\small ABLE}~3. (continued)

\bigskip

\begin{tabular}{rrrcccccccl}
 \tableline
 No. & \multicolumn{2}{c}{Position}                 & Size & $V$ & $I$ &($V-I$)
 & Morp.  & D80          & Radial   & Notes \\
     & \multicolumn{2}{c}{$\alpha (1950.0) \delta$} &      &     &     &      
 &        & No.          & velocity &       \\
 (1) & (2) \ \ \                            & (3) \ \        &
 (4)    & (5) & (6) & (7) & (8) & (9) & (10)          & \ \ (11) \\ 
 \tableline
     & $^{\rm h}$ \ $^{\rm m}$ \ $^{\rm s}$ & \deg \ \ $'$ \ &
 [$''$] &     &     &     &     &     & [km s$^{-1}$] &          \\
 91 & 1\ 12\ 44 &    0\ 06.9 & 10 & 18.3 & 17.1 & 1.2 &?     &     &       & \\
 92 & 1\ 12\ 44 &    0\ 07.9 &  9 & 18.5 & 17.6 & 0.9 &S$e$? &     &       &
      b \\
 93 & 1\ 12\ 45 & $-$0\ 01.0 &  8 & 18.7 & 17.7 & 1.0 &S$e$  &     &       &
      sb \\
 94 & 1\ 12\ 45 &    0\ 00.1 &  6 & 18.6 & 17.4 & 1.1 &S$e$  &     &       &
      COM ? \\
 95 & 1\ 12\ 46 & $-$0\ 01.9 & 13 & 17.4 & 16.2 & 1.3 &S$e$  &     &       &
      dist \\
 96 & 1\ 12\ 47 & $-$0\ 00.3 & 40 & 15.5 & 14.2 & 1.2 &S$e$  &  59 & 13074 &
      COM \\
 97 & 1\ 12\ 48 &    0\ 04.6 & 16 & 17.1 & 16.0 & 1.0 &E     &     &       &
      sb \\
 98 & 1\ 12\ 48 &    0\ 10.5 &  8 & 18.7 & 17.4 & 1.3 &S$l$  &     &       & \\
 99 & 1\ 12\ 48 &    0\ 11.1 &  6 & 18.6 & 17.9 & 0.7 &?     &     &       &
      b \\
100 & 1\ 12\ 50 & $-$0\ 09.1 &  7 & 18.8 & 17.7 & 1.2 &?     &     &       & \\
101 & 1\ 12\ 51 & $-$0\ 07.5 &  6 & 19.0 & 18.0 & 1.0 &S$e$  &     &       &
      sb \\
102 & 1\ 12\ 51 & $-$0\ 00.7 & 23 & 16.8 & 15.6 & 1.2 &S$e$  &  58 &       & \\
103 & 1\ 12\ 51 &    0\ 01.1 & 23 & 16.0 & 14.7 & 1.3 &E     &  57 & 14083 & \\
104 & 1\ 12\ 52 & $-$0\ 11.7 & 19 & 16.6 & 15.4 & 1.2 &S0    &  27 &       & \\
105 & 1\ 12\ 52 &    0\ 02.8 &  7 & 15.8 & 14.9 & 0.8 &?     &     &       &
      DBL, b \\
106 & 1\ 12\ 52 &    0\ 07.6 & 10 & 18.4 & 17.2 & 1.2 &S0/S$e$&    &       & \\
107 & 1\ 12\ 53 & $-$0\ 03.5 & 15 & 16.8 & 15.8 & 1.1 &S0/S$e$& 35 &       &
      sb \\
108 & 1\ 12\ 53 & $-$0\ 03.0 & 12 & 17.4 & 16.1 & 1.3 &E     &  36 &       & \\
109 & 1\ 12\ 54 &    0\ 03.6 &  7 & 18.6 & 17.7 & 0.8 &S$e$? &     &       &
      LSB, b \\
110 & 1\ 12\ 56 & $-$0\ 05.6 & 16 & 17.4 & 16.1 & 1.3 &S$e$  &  34 &       &
      COM, b-comp, dist \\
111 & 1\ 12\ 56 &    0\ 04.0 & 11 & 17.9 & 16.8 & 1.1 &E/S0  &     &       & \\
112 & 1\ 12\ 57 & $-$0\ 02.4 & 26 & 15.6 & 14.4 & 1.2 &E     &  56 &       & \\
113 & 1\ 12\ 57 &    0\ 02.1 & 27 & 15.4 & 14.1 & 1.3 &E     &  55 & 13944 &
      COM \\
114 & 1\ 12\ 57 &    0\ 02.9 & 15 & 17.1 & 15.8 & 1.3 &E/S0  &     &       &
      DBL \\
115 & 1\ 12\ 58 &    0\ 01.5 & 26 & 15.8 & 14.6 & 1.2 &E     &  54 & 14008 &
      DBL \\
116 & 1\ 12\ 59 & $-$0\ 11.6 & 13 & 17.7 & 16.5 & 1.2 &S$e$  &     &       & \\
117 & 1\ 13\ 02 &    0\ 07.7 & 20 & 17.3 & 16.2 & 1.1 &S0/S$e$&    &       &
      dist \\
118 & 1\ 13\ 02 &    0\ 10.6 &  8 & 18.7 & 17.5 & 1.1 &?     &     &       & \\
119 & 1\ 13\ 04 & $-$0\ 10.4 & 12 & 18.0 & 16.9 & 1.1 &S$e$  &     &       & \\
120 & 1\ 13\ 04 & $-$0\ 07.4 & 13 & 18.2 & 17.1 & 1.0 &S$e$  &     &       &
      LSB, DBL, dist, sb \\
 \tableline
\end{tabular}
\end{center}
\end{table*}

\clearpage

\begin{table*}
\begin{center}

T{\small ABLE}~3. (continued)

\bigskip

\begin{tabular}{rrrcccccccl}
 \tableline
 No. & \multicolumn{2}{c}{Position}                 & Size & $V$ & $I$ &($V-I$)
 & Morp.  & D80          & Radial   & Notes \\
     & \multicolumn{2}{c}{$\alpha (1950.0) \delta$} &      &     &     &      
 &        & No.          & velocity &       \\
 (1) & (2) \ \ \                            & (3) \ \        &
 (4)    & (5) & (6) & (7) & (8) & (9) & (10)          & \ \ (11) \\ 
 \tableline
     & $^{\rm h}$ \ $^{\rm m}$ \ $^{\rm s}$ & \deg \ \ $'$ \ &
 [$''$] &     &     &     &     &     & [km s$^{-1}$] &          \\
121 & 1\ 13\ 04 & $-$0\ 02.8 & 10 & 18.3 & 17.1 & 1.2 &S0/S$e$&    &       & \\
122 & 1\ 13\ 04 &    0\ 06.0 &  8 & 18.5 & 17.6 & 1.0 &S$e$  &     &       &
      sb \\
123 & 1\ 13\ 06 & $-$0\ 09.8 & 20 & 16.6 & 15.4 & 1.2 &E     &  26 &       & \\
124 & 1\ 13\ 06 & $-$0\ 04.5 & 19 & 17.1 & 15.9 & 1.2 &S$e$  &     &       & \\
125 & 1\ 13\ 06 &    0\ 04.9 &  5 & 19.0 & 18.0 & 1.1 &?     &     &       &
      LSB \\
126 & 1\ 13\ 07 & $-$0\ 10.6 &  7 & 18.4 & 17.2 & 1.2 &S$l$  &     &       & \\
127 & 1\ 13\ 09 &    0\ 04.4 &  8 & 18.4 & 16.5 & 1.9 &E     &     &       & \\
128 & 1\ 13\ 09 &    0\ 06.0 & 10 & 18.5 & 17.1 & 1.4 &S$e$  &     &       & \\
129 & 1\ 13\ 11 & $-$0\ 01.8 & 28 & 16.4 & 15.1 & 1.3 &S$e$  &  53 & 13436 & 
      COM ? \\
130 & 1\ 13\ 11 & $-$0\ 00.9 & 17 & 16.8 & 15.6 & 1.2 &S$e$  &  52 & 13044 & \\
131 & 1\ 13\ 11 &    0\ 04.1 & 23 & 16.1 & 15.1 & 1.0 &S$e$  &  78 &       &
      b \\
132 & 1\ 13\ 14 & $-$0\ 01.3 & 26 & 15.9 & 14.8 & 1.1 &SB$e$ &  51 &       &
      dist, sb, b-arm \\
133 & 1\ 13\ 14 &    0\ 08.7 & 11 & 17.9 & 16.8 & 1.1 &S$e$  &     &       & \\
134 & 1\ 13\ 15 & $-$0\ 11.6 &  8 & 18.6 & 17.7 & 0.9 &S$e$  &     &       &
      dist, b \\
135 & 1\ 13\ 16 & $-$0\ 09.0 & 13 & 16.0 & 14.6 & 1.4 &E     &     &       & \\
136 & 1\ 13\ 18 & $-$0\ 08.2 &  7 & 19.1 & 18.0 & 1.1 &S$e$  &     &       & \\
137 & 1\ 13\ 18 & $-$0\ 06.9 & 29 & 16.3 & 14.8 & 1.5 &S$l$  &  33 &       &
      COM, r-comp \\
138 & 1\ 13\ 18 &    0\ 00.8 &  7 & 19.4 & 18.4 & 1.0 &S$e$  &     &       &
      sb \\
139 & 1\ 13\ 20 &    0\ 06.7 & 12 & 17.9 & 16.9 & 0.9 &S$e$  &     &       &
      b \\
140 & 1\ 13\ 20 &    0\ 08.2 &  6 & 19.3 & 18.4 & 0.8 &S$l$/I&     &       &
      b \\
141 & 1\ 13\ 22 & $-$0\ 11.2 &  9 & 17.1 & 16.3 & 0.8 &S$e$? &22,23&       &
      DBL, b \\
142 & 1\ 13\ 23 & $-$0\ 09.9 & 12 & 17.7 & 16.4 & 1.3 &E     &     &       & \\
143 & 1\ 13\ 23 &    0\ 02.2 & 10 & 18.0 & 16.7 & 1.3 &E     &     &       & \\
 \tableline
 \tableline
\end{tabular}
\end{center}
\end{table*}

\clearpage

\begin{table*}
\begin{center}

Notes to T{\small ABLE}~3

\end{center}

Columns:

(1) ID-number

(2), (3) Position of the galaxy center in 1950.0 equinox.
The position accuracy is about 5~arcsec.

(4) Major axis diameter of the galaxies at 23~mag~arcsec$^{2}$
isophote in $I$ band.

(5), (6), (7) $V$, $I$, and $(V-I)$ mag.
Tabulated $(V-I)$ colors do not always coincide with $V$ magnitude minus
$I$ magnitude because of the round-off error. 

(8) Eye inspected morphology; category is either of
E, S0, early S [Sa - Sb] (S$e$), late S [Sc - Sd] (S$l$), I.
If we can see any bar structure, we add `B'.

(9) The ID-number given in Dressler (1980).

(10) Measured heliocentric radial velocities.

(11) Some notes.
We use the following abbreviations;
LSB: low surface brightness;
DBL: double system (the companion has the comparable size);
COM: with a small companion
(the number in parenthesis is the ID-number of the companion,
if it is cataloged as another object in this list);
comp: companion;
dist: distortion;
r: red;
b: blue;
sb: semi-blue;
swb: somewhat blue.
No. 39 is the only $IRAS$ source and No. 81 is the only radio source; see text.

\end{table*}

\clearpage

\begin{table*}
\begin{center}

T{\small ABLE}~4. Distribution of the morphology.

\bigskip

\begin{tabular}{clccccc}
 \tableline
 \tableline
 Morphology\tablenotemark{a} & Number  &
 Fraction (\%)\tablenotemark{b} &
 \multicolumn{2}{c}{Blue galaxies} &
 \multicolumn{2}{c}{Blue and semi-blue galaxies} \\
 & & & number & fraction\tablenotemark{c} &
 number &
 fraction\tablenotemark{c} \\
 \tableline
 E                &  \ 32.5 &  \ 25 & \ 0 & \ 0 & \ 1 & \ 3 \\
 S0               &  \ 12   & \ \ 9 & \ 0 & \ 0 & \ 2 &  17 \\
 S$e$ + SB$e$     &  \ 74.5 &  \ 58 &  15 &  20 &  29 &  39 \\
 S$l$ + SB$l$     & \ \ 6.5 & \ \ 5 & \ 3 &  46 & \ 3 &  46 \\
 I                & \ \ 3.5 & \ \ 3 & \ 2 &  57 & \ 2 &  57 \\
 ?                &  \ 14   &       & \ 3 &  21 & \ 3 &  21 \\
 \tableline
 Total            &   143   &   100 &  23 &  16 &  40 &  28 \\
 \tableline
 \tableline 
\end{tabular}

\bigskip

Notes to T{\small ABLE}~4

\end{center}

$^{\rm a}$
Barred and non-barred are combined together.
Objects with uncertain assignments are also included.
For objects assigned to two categories,
we count a half of that in each category.

$^{\rm b}$
Fraction is that to the total number of objects
with assigned morphologies, 129.

$^{\rm c}$
Fraction is that to the total number in each morphological type.

\end{table*}